\documentclass[%
 reprint,
superscriptaddress,
 amsmath,amssymb,
 aps,
longbibliography
]{revtex4-2}

\usepackage{graphicx}
\usepackage{dcolumn}
\usepackage{bm}
\usepackage{upgreek} 

\usepackage[colorlinks=true, allcolors=blue]{hyperref}
\begin{document}

\title{A decay microscope for trapped neon isotopes}

\author{Ben Ohayon}
\email{benohayon@gmail.com}
\affiliation{
 Racah Institute of Physics, Hebrew University, Jerusalem 91904, Israel
}
\affiliation{
Institute for Particle Physics and Astrophysics, ETH Z\"urich, CH-8093 Z\"urich, Switzerland. 
}

\author{Hitesh Rahangdale}%

\author{Elad Parnes}

\affiliation{
 Racah Institute of Physics, Hebrew University, Jerusalem 91904, Israel
}
\author{Gedalia Perelman}
\author{Oded Heber}
\affiliation{
The Weizmann Institute of Science, Rehovot 76100, Israel
}
\author{Guy Ron}
\affiliation{
 Racah Institute of Physics, Hebrew University, Jerusalem 91904, Israel
}
\date{\today}

\begin{abstract}

We review the design, simulation, and tests, of a detection system for measuring the energy distribution of daughter nuclei recoiling from the beta-decay of laser trapped neon isotopes. 
This distribution is sensitive to several new physics effects in the weak sector.
Our `decay microscope' relies on imaging the velocity distribution of high energy recoil ions in coincidence with electrons shaken-off in the decay.
We demonstrate by way of Monte-Carlo simulation, that the nuclear microscope increases the statistical sensitivity of kinematic measurements to the underlying energy distribution, and limits the main systematic bias caused by discrepancy in the trap position along the detection axis.

\end{abstract}

\maketitle


\section{\label{sec:Intro}Introduction}

The last decade witnessed a plethora of applications of experimental methods from the field of atomic molecular and optical (AMO) physics to low-energy nuclear physics observables, with motivations spanning nuclear-structure \cite{2007-Blaum,2016-Struc}, nuclear-astrophysics \cite{2004-NucAstEIBT,2013-MassAstro}, and searches for physics beyond the standard model (BSM; \cite{2018-Review,2019-gazit}).

The \textit{static} (ground or long lived isomeric) properties of stable and short-lived nuclei, are especially useful for such investigations.
The isotopic masses, which grant access to Q-values and separation energies, are routinely extracted using ion traps \cite{2013-Kluge}, and multiple reflection mass-spectrometry \cite{2013-MRTOF};
and the long interrogation times in a magneto-optical-trap (MOT) enable to identify trace amounts of isotopes for radiometric dating \cite{1999-ATTA, 2003-ATTA},
and preform laser spectroscopy of ultracold short-lived samples, which is useful for studying nuclear structure \cite{2007-8He, 2018-Cs, 2018-FrIS}, measurement of atomic parity violation \cite{2013-FRAPV}, and searches for a permanent electron electric dipole moment (EDM; \cite{2007-RaAPV, 2016-FrCYRIC,2017-FrCYRIC,2016-225RA}).

For short lived isotopes, which decay predominantly within the small trap volume, the collision-free environment enables the measurement of \textit{dynamic} decay properties such as branching ratios \cite{2018BranchIon, 2019-BR-MRTOF}, waiting times \cite{2010-FAIR, 2013-BetaDelay}, and angular correlations of decay products. These \textit{angular correlations} are used in the search for BSM physics in the weak sector \cite{2009behr, 2014-BehrTraps}. In fact, determinations of correlations to the level of few per mil are competitive and complementary with high-energy searches \cite{2013-LHCera, 2018-Review}.

Advancements in AMO physics, and in the production, separation and transport of rare radioactive isotopes, enable an increasing variety of isotopes to be trapped in a MOT. See table \ref{tab:traps} for an overview of experiments which are ongoing, recently commissioned or in preparation, and \cite{1997-MOTr,2000-MOTr,2003-MOTr,2014-BehrTraps} for reviews of past experiments.

In this paper we detail the neon atom trap experiment (NeAT), focusing on the detection system for charged particles recoiling from the trap. It relies on our recent merging of advanced ion-imaging techniques with the MOT \cite{2019-MOTVMI}, and an adaptation of the setup to the high energies encountered in nuclear decay, which is detailed here. We demonstrate, through Monte-Carlo (MC) simulations, that this \textit{decay microscope} is able to determine angular correlations and branching ratios at the few pro-mil level by collecting a few $10^7$ coincidence events, while removing the main sources of noise, and systematic uncertainties associated with the trap size and position.

Considering ongoing and planned experiments in the field, we identify that the main impact of angular-correlation and branching-ratio measurements utilizing several neon isotopes, is in constraining BSM tensor interactions coupled to right-handed neutrinos, and extracting the Cabibbo-Kobayashi-Maskawa (CKM) matrix element for superallowed Fermi and mirror and transitions.

\begin{table}[!htbp]
  \centering
\caption{Overview of rare radioactive atom trap experiments, focusing on ones recently conducted and those that are in preparation. ATTA - Atom Trap Trace Analysis, LS - Laser spectroscopy, APV - Atomic parity violation, EDM - electric dipole moment.}
  
 \begin{ruledtabular} 
\begin{tabular}{lllll}

Isotope    &  Half-life   & Experiment & Facility & Ref. \\
\cline{1-5} \\

$^{6}$He  & $0.8$ s & Decay         & UW/ANL   & \cite{2016-HongMCP,2016-HongTh,2017HongCharge,2019-6HeThesis} \\

$^{18,19,23}$Ne &  $1.7-37$ s & Decay          & SARAF   & \cite{2018-WeakSARAF,2018-EPJA} \\

$^{35}$Ar  & $1.8$ s         & Decay     & Leuven    & \cite{2015-Glover}   \\

$^{39}$Ar  & $269$ y         & ATTA        & Heidelberg &\cite{2014-39Ar,2018-39Ar}    \\

$^{37}$K  & $1.2$ s & Decay & TRIUMF        & \cite{2018-37K}    \\

$^{38m}$K  & $0.9$ s & Decay & TRIUMF        & \cite{2018-Review}    \\

$^{81}$Kr  & $2.2\times10^5$ y         & ATTA        & ANL,USTC &  \cite{2012-ATTA3,2015-ATTAUSTC}  \\

$^{85}$Kr  & $11$ y         & ATTA        & Columbia & \cite{2013-85Kr,2018-85Kr}
\\ & & & ANL, USTC & \cite{2012-ATTA3, 2015-ATTAUSTC}    \\

$^{92}$Rb  & $4.5$ s   & Decay      & TRIUMF        & \cite{2018-92Rb}    \\

$^{131}$Cs  & $9.7$ d   & Decay  & UCLA & \cite{2019-131Cs,HUNTER}  \\

$^{134-142}$Cs  & diverse   & $\gamma$-Laser, LS  &  Jyvaskyla &  \cite{2018-CSG, 2018-Cs}  \\

$^{208-211,213}$Fr  & $35-191$ s   & LS, APV  &  TRIUMF &  \cite{2013-FrTRIUMF, 2018-FrIS}  \\
$^{210}$Fr  & $3.2$ m   & EDM     & CYRIC   & \cite{2016-FrCYRIC,2017-FrCYRIC}   \\

$^{225}$Ra  & $15$ d   & EDM    & ANL  & \cite{2016-225RA,2019-225Ra}  \\

\end{tabular}%
\end{ruledtabular}
\label{tab:traps}%
\end{table}%


Assuming a general, Lorentz-invariant, interaction Hamiltonian density \cite{1956-LeeYang}, 
the derived differential decay rate $\Gamma$, of non-oriented nuclei reads, \cite{1957-Jackson}
\begin{equation}\label{eq:beta}
    \frac{d^3\Gamma}{dE_\beta d\Omega_\beta d\Omega_\nu} \propto 
    1+a\frac{\boldsymbol{p}_\beta\cdot \boldsymbol{p}_\nu}{E_\beta E_\nu}+
    b\frac{m_\beta}{E_\beta},
\end{equation}
where we denote three-momenta as $\boldsymbol{p}$, scalar energy as $E$, rest mass $m$, and do not include various corrections and scale factors.

Observing eq. \ref{eq:beta}, there are two coefficients which govern the kinematic observables of the decay products:

\begin{enumerate}
\item The $\beta$-$\nu$ angular coefficient 
\begin{equation}\label{eq:a}
\begin{split}
    a\propto
             &~|M_F|^2~(|C_V|^2+|C'_V|^2-|C_S|^2-|C'_S|^2)\\
-\frac{1}{3} &~|M_{GT}|^2~(|C_A|^2+|C'_A|^2-|C_T|^2-|C'_T|^2)
\end{split}
\end{equation}

\item The Fierz interference term,
\begin{equation}
\begin{split}
b\propto
|M_F|^2  ~(C_S+C'_S)+|M_{GT}|^2(C_T +C'_T),
\end{split}
\label{eq:b}
\end{equation}
\end{enumerate}
where we denote the relevant Lee-Yang parameters $C_i^{(')}$ assuming negligible time violation (see \cite{2018-Review} for a more general derivation).
$M_F$ denotes the Fermi matrix element, and $M_{GT}$ the Gamow-Teller element, which depend on the nuclear structure of the isotope in question.

To determine the statistical sensitivity of direct recoil ion energy measurements to $a$ for various conditions, we preform a Monte-Carlo (MC) simulation. We fit the recoil ion energy distribution resulting from equation \ref{eq:beta}, given in \cite{2016-FierzOscar}, by maximum likelihood estimation, leaving $a$ a free parameter, and extract the standard deviation of the extracted values from the fit. Our results are given in table \ref{tab:Sens1}. Due to the quadratic dependence of $a-a_{sm}$ on new physics, we assume $b=0$ in the fitting process, an assumption which is aligned with the extraction of $\rho$ for mixed transitions, and a search for new physics coupled to right-handed neutrinos for pure ones. 
We refer the reader to \cite{2016-FierzOscar} for the sensitivity of the spectrum to $b$ assuming $a=0$.
Pure transitions are simulated with an endpoint energy of $4$ MeV, and the results are comparable in the range $0.5-4$ MeV \cite{2016-FierzOscar}. As a realistic scenario, and in accordance with ongoing and planned experiments, we assume $10^7$ detected events for recoil ion detection. 

\begin{table}[!htbp]
  \centering
  \caption{
  Sensitivity at the 1$\sigma$ level, of recoil ion  spectral measurements, with $10^7$ events. $S(a)=\sigma(a)/a, S(BR)=\sigma(BR)/BR$ pertain to angular correlation and branching ratios respectively, and given in percentages. Subscripts: F - Pure Fermi, GT - Pure Gamow-Teller, and mass numbers for neon isotopes described in the text.
The bottom row lists the most precise determinations from the literature.
}
  
 \begin{ruledtabular} 
\begin{tabular}{lllllll}

   $S(a_F)$ & $S(a_{GT})$  &  $S(BR_{18})$&  $S(a_{19})$ &  $S(a_{23})$ &$S(BR_{23})$\\\\

  $0.07$  & $0.19$&  $0.2$ &$1.1$ &$0.24$ & $0.3$\\
\\

$0.5$ \cite{2005-BehrK} & 
$0.9$ \cite{1963-Johnson} & $2.7$ \cite{1975-18NeBR} &$3.6$ \cite{1975-19NE} &$9.1$ \cite{1963-carlson} &$9.4$ \cite{1957-Penning}

\end{tabular}%
\end{ruledtabular}
\label{tab:Sens1}%
\end{table}%

\section{Opportunities with Neon isotopes}\label{chap:iso}

The inert nature of noble gases makes it possible to extract them from a solid target through diffusion while leaving most contaminants confined within the matrix.
Such a scheme dispenses with the need for an extensive ionization and mass separation apparatus.
Noble gases are also trappable by laser radiation through driving of a dipole transition from a long-lived atomic metastable state \cite{2012-BirklReview}. For neon, a $640$ nm laser drives the $^3$P$_2$-$^3$D$_3$ closed transition, the metastable lifetime is $15$ s \cite{2003-NeLifetime,2011-Feld,2012-Li}, and the isotope shifts are of order $2$ GHz \cite{2019-IS}.

Neon possesses three radioisotopes having appropriate half-lives. Longer than the extraction time from the target, which is of order of few hundred milliseconds; and shorter than, or comparable to, the intrinsic lifetime of the excited atomic state, which is extended up to $30$ s through saturation of the cooling transition \cite{2018-WeakSARAF}. These isotopes are $^{18,19,23}$Ne, which display diverse modes of decay, and so offer a broad experimental campaign using a single trapping apparatus.

Due to its long half-life of $37.15(3)$ s \cite{2015-HL18NE}, and low threshold for production by neutrons ($3.8$ MeV), our first campaign focuses on $^{23}$Ne, which decays by $\beta^-$ emission to $^{23}$Na. 
The decay is pure GT, up to an isospin-forbidden correction of order $0.1\%$ for the first excited branch \cite{1958-Ridley}, and within the SM, a precision correlation measurement will test this correction.
The previous most precise correlation measurement in $^{23}$Ne yielded $a=-0.33(3)$ \cite{1963-carlson}, with fractional uncertainty limited to $9\%$ by the uncertainty of the absolute branching ratio (BR) of $32(3)\%$ to the first excited state as measured by \cite{1957-Penning}. Thus, a better determination of the BR is essential in order to make a substantial improvement. A dedicated campaign in our group aims at $1\%$ relative uncertainty \cite{2019-BR}, however, our MC simulation shows that a uncertainty of that order still contributes $0.75\%$ to the relative uncertainty in $a$. Since the BR and correlations affect the ion energy distribution differently, it makes sense to use a new, $1\%$ determination of the BR as a prior, and deduce a more stringent BR from the recoil energy distribution itself. Such an approach was recently and successfully utilized in beta-delayed neutron spectroscopy \cite{2013-BetaDelay, 2014-BetaDelay,2019-BetaDelay}.
We find that for $10^7$ $^{23}$Ne simulated decay events, a two parameter fit extracts the branching ratio with a maximal relative bias of $0.3\%$, does not introduce significant systematic shift in $a$, and only slightly increases its statistical uncertainty, from $0.18\%$ to $0.24\%$.
Observing table \ref{tab:Sens1}, and provided that systematics and theoretical corrections are under control \cite{2019-gazit}, the detection of $10^6$ recoil ions from $^{23}$Ne decay already improves on the state of the art. Nevertheless, to be competitive with ongoing and planned campaigns involving pure GT branches \cite{2017HongCharge,2016-HongTh,2018-Ish,2019-BPT,2019-32Ar,2019-WISARD}, roughly $10^7$ events are desired.

$^{18}$Ne decays mainly to the ground state of $^{18}$F through a pure GT transition, and to an excited state by a superallowed Fermi transition. The possibility to extract $V_{ud}$ from the Fermi branch motivated precise determinations of its half-life \cite{2013-18Ne,2015-HL18NE}, and is currently limited by the knowledge of the branching ratio, of which the most precise published value is $7.7(2)\%$ \cite{1975-18NeBR}.
Whereas conventional counting experiments for determining this BR suffer from the build-up of unstable $^{18}$F, with half-life of roughly two hours, creating a time-dependant background; an atomic trap is isotope-selective, and upon their creation, $^{18}$F ion are evacuated by the vacuum-system or adhere to the chamber walls within milliseconds, without contributing to the detected recoil ions emanating from the trap center.
Assuming standard model correlations, a well-motivated scenario for CKM determination, a one parameter fit to the distribution of energy of $10^7$ recoil ions determines the branching-ratio to $0.2\%$. Under the reasonable assumption that the Q-value is improved soon at ISOLTRAP \cite{2018-18NeQ}, this BR will allow the $\mathcal{F}t$ value to be determined to $0.2\%$. The importance of a high precision $\mathcal{F}t$ value for $^{18}$Ne lies in the fact that it is the isotope most sensitive to the differences between the predictions of various models for nuclear-structure-dependant corrections \cite{2015-Ne18,2015-18NeThesis}, whose uncertainty now dominates the $\mathcal{F}t$ values for the most precise superallowed decays \cite{2016-C10_bF}.

$^{19}$Ne decays almost exclusively to the ground state of $^{19}$F through a mixed $T=1/2\rightarrow1/2$ mirror transition. Its mixing ratio $\rho=-1.5995(45)$, is close to the minimum value of $-\sqrt{3}$ for such transitions, and so the high precision it enjoys stems from a modest precision in the determination of correlation coefficients, Namely $A=-0.0391(14)$ \cite{1975-19NE}. Its singularly well-determined mixing ratio, makes $^{19}$Ne the best candidate (along with $^{37}$K) for determining $V_{ud}$ in mirror transitions. This fact motivated very precise measurements of the $^{19}$Ne decay observables, namely its Q-value \cite{2008-Geitner,2017ame2016}, half-life \cite{2012-19NeHL,2014-19Ne,2017-19Ne_HL}, and branching ratios \cite{1993-19NeBR,2019-19Ne_BR}; as well as recent calculations of the relevant radiative and isospin-breaking corrections \cite{2019-MirrorRad,2019-IS_Mirror}. 
Calculating the beta-neutrino correlation coefficient from the mixing ratio, we arrive at $a_{19}=0.0414(15)$, having a $4\%$ relative uncertainty. Observing table \ref{tab:Sens1}, accumulation of $10^6$ recoil ion events in the decay of $^{19}$Ne already improves on the state of the art, and reduces the uncertainty in the prediction of $A$, which is well-motivated considering the discrepancies in other determinations of $A$, as well as deviations from the conserved-vector-current hypothesis as discussed in \cite{2019-19Ne_BR}. To complement the highly precise determination of $V_{ud}$ from the body of $0^+\rightarrow0^+$ decay data \cite{2015-SuperAllowed}, and considering ongoing correlation measurement efforts with nuclear mirror  \cite{2011-betanuMirror,2015-Ganil,2016-VudSt.,2015-Glover,2018-37K,2018-GanilShake,2019-LPC,2019-St.}, and neutron \cite{2009-Nab,2017-aCORN,2018-aspect,2019-Nab,2019-spect, NoMoS} decays, roughly $10^7$ detected events are desired, especially in light of the emergence of a significant deviation from CKM unitarity obtained in some of the recent analyses \cite{2019-CKM,2019-CKMu,2019-SengCKM,2019-CMK2}.

See \cite{2018-EPJA} for the proposed production scheme for Neon isotopes at SARAF-II.

\section{Detection System}

The statistical sensitivity described in table \ref{tab:Sens1} and chapter \ref{chap:iso} pertains to a direct detection of the daughter ion recoil energies, which are of order of few hundred eVs, and so can not be measured precisely by direct calorimetric means. In cases where the daughter nuclei are short-lived, one may infer this distribution from correlation measurements in beta-delayed proton \cite{2019-TAMUTRAP,2019-WISARD} or alpha \cite{2015-Li8,2019-BPT} emitters. For correlation measurements in isotopes such as neon, which decay to a stable or long-lived daughter, one infers the energy from kinematic observables such as time-of-flight (TOF) and hit position using charged particle detectors.

Systematic uncertainties, which usually dominate in this sort of research, are dependent upon the details of the implemented detection scheme, as well as contributions from the background. The main source of background is decay products from neon in the atomic ground state, which does not interact with the trapping light and may decay close to the source volume before being pumped out of the system.
\subsection{Deflection and compression}
\begin{figure}[!b]
\centering
\includegraphics[width=0.9\columnwidth,trim={2 0 8 5},clip]{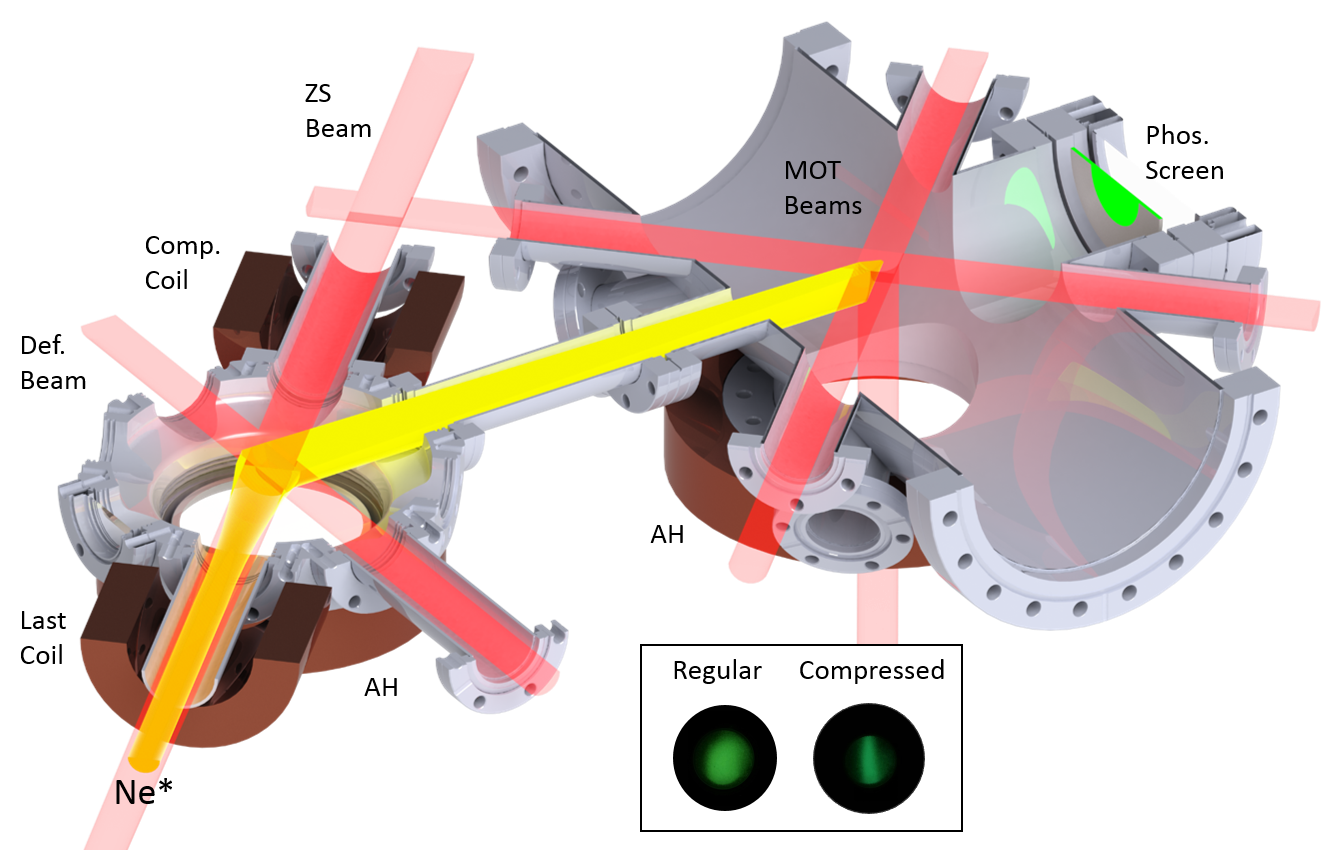}
\caption{Deflection CAD model. MCP phosphor-screen image with regular and 1D compressed beam is shown in the inset. For ion-imaging electrodes, see Fig. \ref{Fig:Elec}. AH - Anti-Helmholtz pair, ZS - Zeeman slower.}
\label{Fig:Def}
\end{figure}
In order to significantly reduce the background from, and collisions with, ground state radioactive neon, we employ an isotope-selective, 45 degree, deflection stage between the metastable source and Zeeman-slower beamline, and the science chamber, in which the MOT and detection system reside. 

Our deflection stage utilizes a single $50$ mW retro-reflected, red-detuned, laser beam which acts as a 1D optical molasses. Shutting-off of the deflection beam enables estimation of background contributions to the detectors without changing any trapping parameters.
The last coil of our segmented Zeeman-slower \cite{2013-ZS, 2015-ZS}, whose field is responsible for the atomic beam velocity, is paired with a compensation coil at the other end of the chamber, which controls the turning position. Adding another coil pair in anti-Helmholz configuration allows for adjusting the strength of the magnetic gradient in the vicinity of the turning point without changing its location (see Fig. \ref{Fig:Def}). The combination of a 1D optical molasses and a magnetic field gradient acts as a cylindrical atomic lens which tilts and focuses the atomic beam in one dimension.
To test and optimize the beam focusing, brightness and angle, we place a microchannel-plate (MCP) detector equipped with a phosphor screen and image the energetic metastable beam for various tuning conditions. With the lens focus unadjusted, the deflected beam diverges and completely covers the detector. By varying the magnetic degrees of freedom, the deflected beam focuses in one dimension, with the unfocused dimension aligned with the vertical trapping beam.
By comparing the loading rate for a MOT placed at the deflection chamber, and one in the science chamber, we estimate the deflection efficiency at $20\%$, which is comparable with other deflection setups relying on push-beams (\cite{2018-37K} - $75\%$, \cite{2016-FrDef} - $50\%$, \cite{2011-Push} - $4\%$ including dipole trap loading), optical-dipole-trap transfer (\cite{2012-ODT} - $25\%$ including dipole trap loading), and a push-and-guide technique for metastable helium (\cite{2019-6HeThesis}, $10-20\%$).

In addition to reducing the ground state neon background, the science chamber is protected from the strong ionizing VUV radiation and charged particles emanating from the RF source \cite{2015-RFs}. At a pressure of a few $10^{-10}$ Torr the trap lifetime is roughly $20$ s, close to the limit imposed by the lifetime of the meatastable state.

\subsection{Coincidence ion-imaging}

To efficiently collect the decay products emerging from the trap volume, it is surrounded by electrodes creating a static field of few kV/cm. The same field folds the ion trajectories in a non-trivial way, thus limiting the ability to recreate the ion energy distribution from the kinematic observables.

The previous three experiments which measured $a$ using trapped radioisotopes relied mainly on TOF detection by triggering on the beta particles or shake-off electrons \cite{2005-BehrK,2009-80Rb,2008-ShakeoffVetter,2016-HongTh}. In all these experiments, a significant contribution to the systematic uncertainty is the uncertainty in the trap position.

To address the challenges above, we designed a new detection scheme, which relies on state-of-the-art charged-particle imaging techniques, namely Velocity-Map-Imaging (VMI), a well-established, efficient, and simple scheme for imaging electron and ion velocity distributions \cite{1997-VMI}. By placing the interaction region at the focus of an electrostatic lens, VMI employs space-focusing, where, to first order, the trap size and position do not affect the particle hit position and so the deduced velocities.

\begin{figure}[!bp]
\centering
\includegraphics[width=0.95\columnwidth,trim={2 0 8 5},clip]{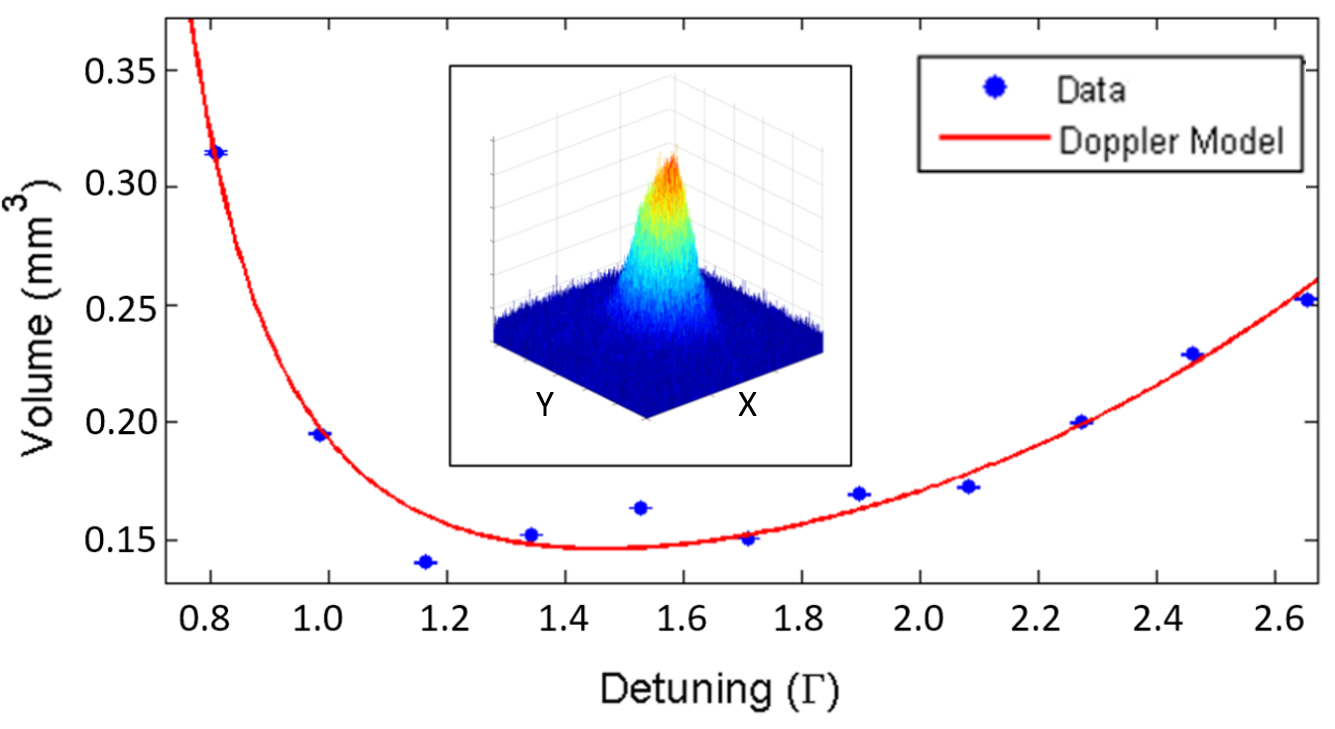}
\caption{Trap volume deduced from CCD camera images (inset). The detuning is in units of the natural linewidth, $\Gamma=8$ MHz, and its effect on volume is well-described by the Doppler model.} 
\label{Fig:Size}
\end{figure}
To considerably reduce background counts, and distinguish between different charge-states of recoiling ions \cite{1963-Carlshake,2012-GANIL6HeC,2017HongCharge,2018-GanilShake,2018BranchIon}, coincidence detection is desired, allowing to measure the recoils TOF in parallel with their hit position. Similarly to \cite{2004-ScielzoShake,2009-80Rb}, we opted to use shake-off electrons as a start signal and detect recoil ions in coincidence.
We tested the first version of the MOT-VMI, including the  coincidence TOF detection, by imaging low-energy ($1$ eV) ions emerging from cold collisions in the a stable neon trap, and deducing branching ratios and penning-ion energy distributions \cite{2019-MOTVMI}.
\begin{figure}[!tp]
\centering
\includegraphics[width=0.9\columnwidth,trim={10 1 6 5},clip]{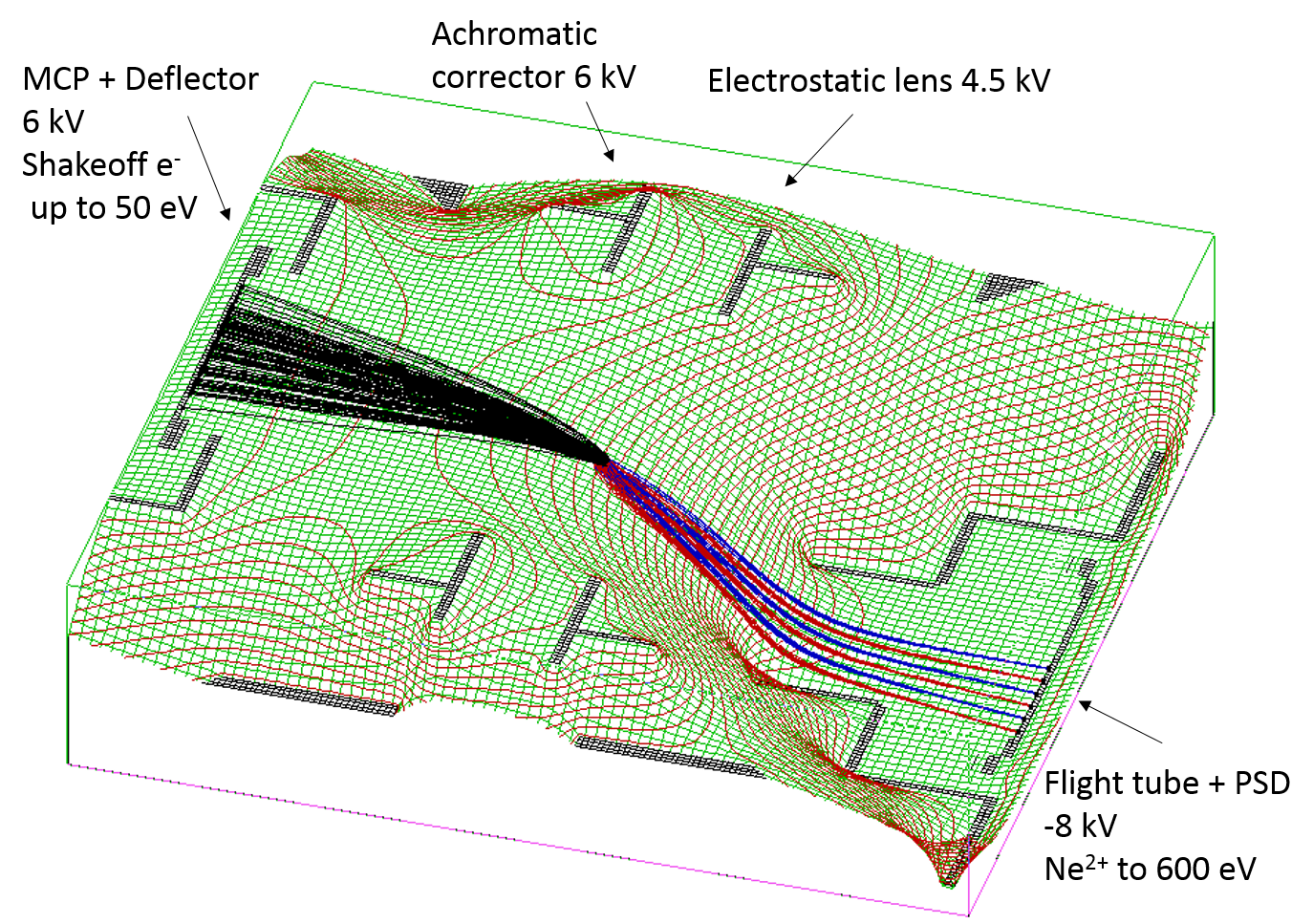}
\includegraphics[width=0.9\columnwidth,trim={10 1 6 5},clip]{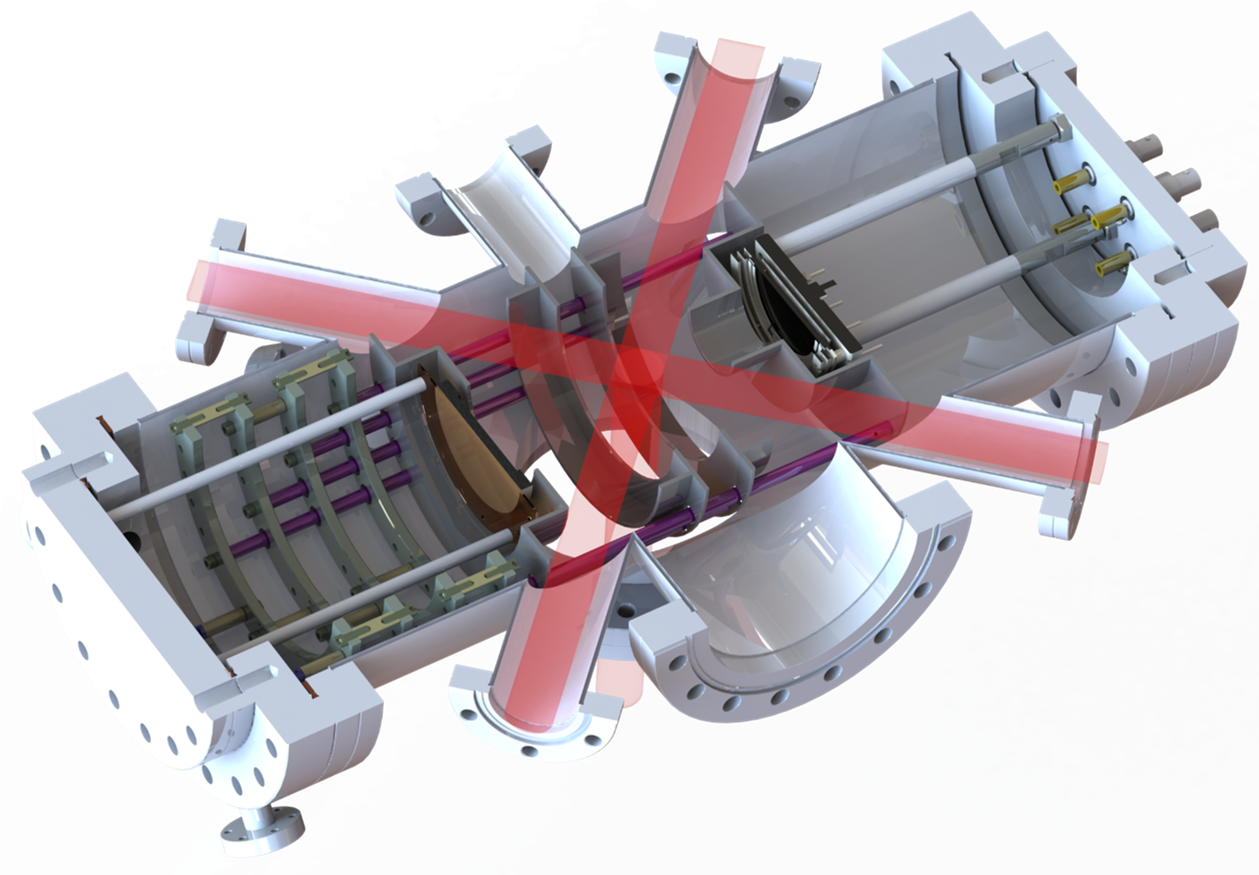}
\caption{
Top: Simion 8.1 simulation of charged particles emerging from the trap volume and focused to the detectors. shake-off electrons up to $50$ eV are collected on the micro-channel-plate (MCP) and give a start signal.
Doubly charged $^{23}$Na ions of various energy groups up to $600$ eV are focused to $0.1$ mm  on a position-sensitive-detector (PSD).
Bottom: CAD model of science chamber including the trapping laser beams, the imaging setup electrodes, and the detectors.
}
\label{Fig:Elec}
\end{figure}

To adapt the MOT-VMI to the higher energy of recoils from nuclear decay (shake-off electrons up to $50$ eV, doubly charged recoil ions up to $600$ eV), whilst keeping the maximal absolute voltages at $8$ kV, we designed the setup with the detectors much closer to the trap volume (see Fig. \ref{Fig:Elec}). The close imaging plane necessitates tuning the lens to have a short focal distance, causing severe chromatic aberrations \cite{2014-HEVMI}, where the focal length is dependant on the ion energy. To correct for these aberrations, while keeping the setup as simple as possible and complying with the trapping lasers restrictions, we added an achromatic corrector lens on the opposite side of the trap. See \cite{2014-HEVMI,2014-Kling,2017-HEVMI} for other high-energy VMI designs.

\begin{figure}[!bp]
\centering
\includegraphics[width=\columnwidth,trim={0 0 0 0},clip]{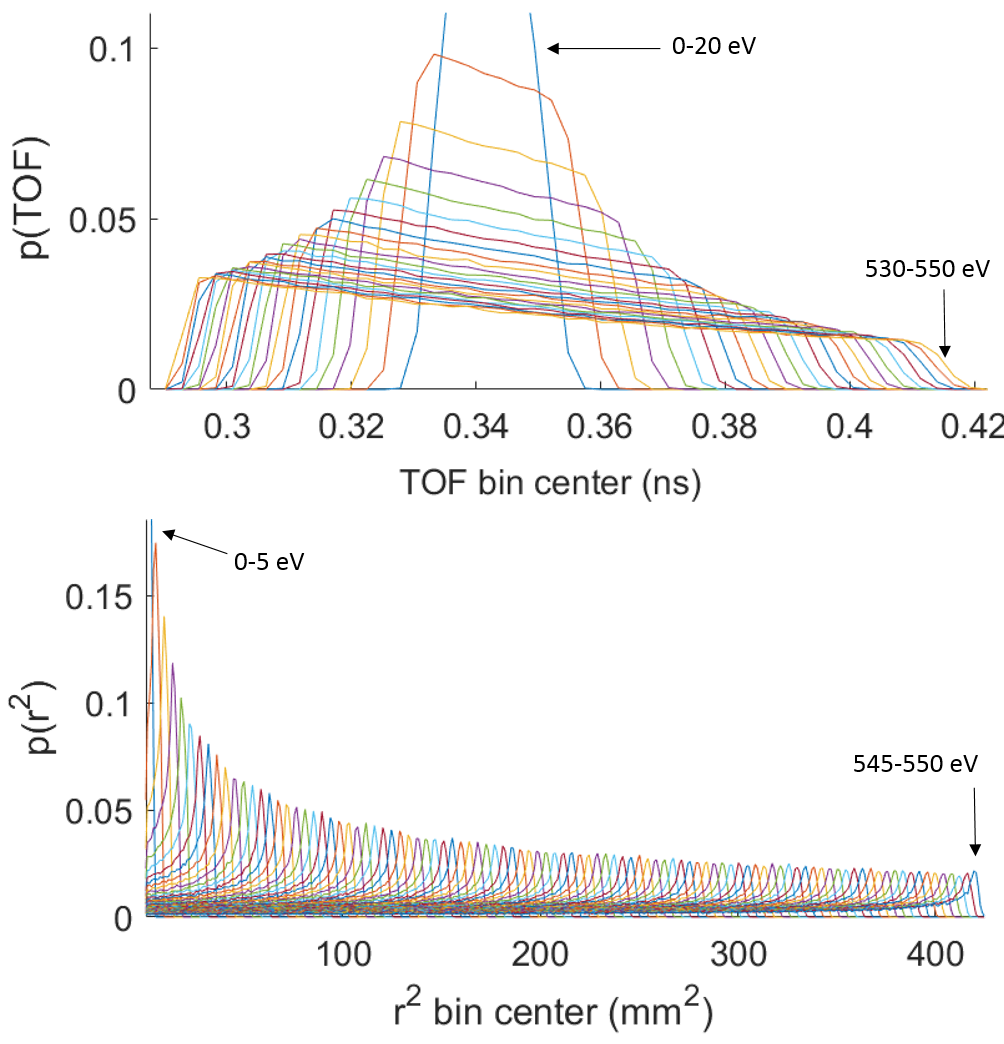}

\caption{Simulated detection system response for different energy groups emitted isotropically from the trap volume. Top: time-of-flight distributions. Bottom: squared hit position distributions. }
\label{Fig:basis}
\end{figure}

The trap size as a function of laser detuning was measured using a calibrated CCD camera and a $^{20}$Ne trap (Fig. \ref{Fig:Size}), and the minimum possible trap radius was determined to be $0.5$ mm (FWHM).
Using the Simion 8.1 software package \cite{2000-Simion}, we optimized the field shape so that each energy group emitted from the decay volume, perpendicular to the detector plane, is focused to a $<0.1$ mm thickness ring on the imaging plane, which is on the order of the resistive-anode-detector inherent resolution (Fig. \ref{Fig:Elec}).
At optimum conditions, the obtained position resolution is $0.5-1\%$, yielding an energy resolution of $1-2\%$. 
To gauge the response of the system, we simulated TOF and squared hit position for recoil ions at various energies, emitted isotropically from the trap volume (Fig. \ref{Fig:basis}).
We find that the TOF resolution is limited by the trap size to $4$ ns, and that the TOF basis functions overlap to a large extent, meaning that the a similar response is obtained for various energy groups, making it difficult to distinguish them. 
Observing the squared hit position basis functions (Fig. \ref{Fig:basis} bottom), each energy group display a sharp peak, corresponding to ions emitted perpendicularly to the detector. The orthogonality of the position basis functions indicate that an energy distribution may be inferred from them with high sensitivity.

\section{Monte-Carlo simulation}

To asses the performance of the entire detection system we conduct a MC simulation, mimicking a complete experiment and its analysis. 
Doubly-charged $^{23}$Na ions are emitted isotropically, with the SM energy distribution for $a=-1/3, b=0$, and from the decay volume of Fig. \ref{Fig:Size}. They are propagated to the detection plane utilizing a SIMION 8.1 simulation of the electrode system of Fig. \ref{Fig:Elec}, and their TOF and hit positions are recorded on an event-by-event basis. A small delay time of $3$ ns caused by the flight time of shake-off electrons to the opposite MCP is subtracted from the results, it has a spread of $0.1$ ns which is negligible as compared with the spread caused by the finite size of the trap.

To deduce the sensitivity to $a$ for TOF and position detection, we compare simulated experimental results with templates generated around the SM prediction, namely $a=-0.2$ and $a=-0.4$, which are propagated through the simulation as well. Even though exotic tensor interactions will cause only $a>-1/3$ (eq. \ref{eq:a}) we find that allowing $a<-1/3$ is crucial for identifying systematic biases caused by discrepancies between the experimental and theoretical fit. 

Incorporating various discrepancies in the MOT size and position between the simulated experimental results and the MC templates, we find that the trap position along the electrode axis causes the largest bias in the results. An unaccounted-for $\pm 0.2$ mm shift in the MOT axial position, which is the limit to our ability of pining-down the trap using TOF of neon dimers, creates a systematic shift of roughly $\mp2\%$ for position detection and $\pm10\%$ for TOF detection. 

To decrease the above bias, which will severely limit a precision determination of $a$, we allow the trap axial position to become a free parameter in the fit, by simulating four fitting templates $Y$, weighted to generate a combined template $T$
\begin{equation}
\begin{split}
T(w_1,w_2)   = w_1  w_2       & Y(a=-0.4,\delta_z=+0.5)\\
     + w_1(1-w_2)             & Y(a=-0.4,\delta_z=-0.5) \\ 
     + (1-  w_1)w_2           & Y(a=-0.2,\delta_z=+0.5)\\
     + (1-  w_1)(1-w_2)       & Y(a=-0.2,\delta_z=-0.5).
\end{split}
    \end{equation}
The axial position and $a$ are deduced by minimizing the negative Log-likelihood function assuming Poisson distribution
\begin{equation}
L(w_1,w_2)=\sum_{bins}{ T(w_1,w_2)-E\cdot Log( T(w_1,w_2) ) },
\label{eq:MLE}
\end{equation}
with $E$ the simulated experimental TOF or hit position distributions, $w_1$ a weight corresponding to $a$, and $w_2$ to the trap position.
The returned correlation coefficient is calculated as $a = -0.4 w_1-0.2 (1-w_1)$, and its uncertainty is derived from the standard deviation $\sigma(a)$, given in table \ref{tab:Sens2}, of the distribution of $a$.

%
Without introducing more degrees of freedom to the fit, we tested for biases caused by other unaccounted-for discrepancies. 
The results, given in table \ref{tab:Sens2} indicate that any discrepancy leads to a large offset using TOF detection, and  a mostly negligible one utilizing position detection with a two-parameter fit. We ascribe this reason to the cancellation of volume effects by the lens system, and to the fact that an axial shift in the trap position behaves to first order as a change in the average field of the imaging system. The largest shift for position detection is caused by an unaccounted-for discrepancy in the lens diameter.
Even though a $\pm0.4$ mm change is partially compensated as a $\mp0.2$ mm offset to the trap axial position, it still leads to $0.4\%$ change in $a/a_{GT}$. It would be thus beneficial to measure the geometry to better accuracy, and to calibrate the magnification of the lens system utilizing the known Q-value for the decay.

\begin{table}[htbp]
  \caption{One standard deviation sensitivity to $a$ assuming $b=0$ obtained by two parameter fitting for $10^7$ events. $z\pm0.2$ denotes the offset in $a/a_{GT}$ caused by a $\pm0.2$ mm discrepancy in the location of the trap along the detection axis, $r+0.2$ the offset by a radial shift of $0.2$ mm from the trap center, and $l\times1.5$ by a $1.5^3$ change to the trap volume. $V_L\pm0.1\%$ denotes a discrepancy in the lens voltage, and $D_L\pm0.4$ denotes $\pm0.4$ mm in the lens diameter.}
  
\begin{ruledtabular} 

\begin{tabular}{lrrrrr}

Method &   $S(a_{GT})$ & $z+0.2$ & $z-0.2$ & $r+0.2$ & $l\times1.5$\\
\cline{1-6} \\

Position & $0.4\%$ & $<0.3$\%  & $<0.2$\%   & $<0.1$\%  & $<0.1$\%   \\
TOF      & $0.8$\% & $3.0\%$   & $1.5\%$  & $2.5\%$ & -$7.5\%$
\\
\cline{1-6} \\

     &   $V_L+0.1\%$ & $V_L-0.1\%$ & $D_L-0.4$ & $D_L+0.4$ \\
\cline{1-6} \\

Position & $<0.2\%$ & $<0.2\%$  & $-0.3\% $   & $ 0.4\%$     \\
TOF      & $1.3$\% & $1.4\%$   & $ 1.1\%$  & $ 1.6\%$ \\

\end{tabular}%
\end{ruledtabular}

\label{tab:Sens2}%
\end{table}%

\section{Summary}

Precise determinations of the energy distribution of recoil ions emerging from nuclear beta-decay enables the determination of branching fractions to various excited states, and the measurement of angular correlations between the decay products.
In light of on-going and planned experiments in this field, the main opportunities to make a significant impact using trapped neon isotopes, is in searching for, or excluding, new tensor physics which is coupled to right-handed neutrinos, as well as extracting the $V_{ud}$ CKM matrix element for mirror and superallowed Fermi transitions.

For correlation measurements in isotopes decaying to a stable or long-lived daughter, it is necessary to deduce energy distributions from the kinematics of the recoil daughter nucleus, necessitating the use of ion or atom traps for confinement and cooling.
In accordance with previous trap experiments, we estimate that our final results will be limited by systematic uncertainties caused by background from decays of non-trapped isotopes, and discrepancies in the determination of the trap position.
To reduce uncertainty associated with background from decay of neon in the ground state, we employ a unique continuous isotope and state-selective deflection which compresses the atomic beam elliptically and displaces the science chamber from the main beamline.

To maximize the sensitivity of a kinematic measurement to the energy distributions, as well as reduce the contribution from the trap size and position on the determined correlation coefficients and branching ratios, we designed a high-energy MOT-VMI, acting as a decay microscope, which images the velocity distribution of recoiling ions from the trap volume, in coincidence with shake-off electrons, enabling TOF determination for charge/mass selection and noise reduction. 
A low-energy version of the MOT-VMI, including the data acquisition logic, timing and trap size characteristics, and background contributions, was thoroughly tested by imaging charged particles emerging from collisions in a dense stable neon trap \cite{2019-MOTVMI}. Here we described the adaptation of the design to the requirements of a decay experiment.
Through extensive MC simulations, we devised a fitting procedure which removes unaccounted-for volume-effects from the determined correlations and enables a significant improvement over the state of the art by collecting a few $10^7$ coincidence events.

\begin{acknowledgments}

The work presented here is supported by grants from the Pazy Foundation (Israel), Israel Science Foundation (grants no. 139/15 and 1446/16), and the European Research Council (grant no. 714118 TRAPLAB). BO is supported by the Ministry of Science and Technology, under the Eshkol Fellowship, and by the Israel Academy of Sciences and Humanities.

\end{acknowledgments}

\bibliography{cleanbib1}

\end{document}